\documentclass[prb,aps,showpacs,prb,amsmath,floatfix,twocolumn]{revtex4}
\usepackage[]{graphicx}
\usepackage{graphicx}
\usepackage{dcolumn}
\usepackage{bm}

\begin{document}
\title{Designer thermal switches: Effect of the contact
material on instantaneous thermoelectric transport
through a strongly interacting quantum dot}
\author{A. Goker$^{1}$ and E. Gedik$^{2}$}

\affiliation{$^1$
Department of Physics, \\
Bilecik University, \\
11210, G$\ddot{u}$l$\ddot{u}$mbe, Bilecik, Turkey
}

\affiliation{$^2$
Department of Physics, \\
Eskisehir Osmangazi University, \\
26480, Meselik, Eskisehir, Turkey
}

\date{\today}

\begin{abstract}
We investigate the effect of contact geometry on
the instantaneous thermoelectric response of a 
quantum dot pushed suddenly into the Kondo regime 
via a gate voltage using time dependent non-crossing 
approximation and linear response Onsager relations.
We utilize graphene and metal contacts for this 
purpose. Instantaneous thermopower displays 
sinusoidal oscillations whose frequency is 
proportional to the energy separation between 
the van Hove singularity in the contact density 
of states and Fermi level for both cases regardless 
of the asymmetry factor at the onset of Kondo 
timescale. The amplitude of the oscillations 
increases with decreasing temperature saturating 
around the Kondo temperature. We also calculate 
the instantaneous figure of merit and show that
the oscillations taking place at temperatures
above the Kondo temperature are enhanced more
than the onesoccurring at lower temperatures
due to the violation of the Wiedemann-Franz law.
Graphene emerges as a more promising electrode 
candidate than ordinary metals in single electron 
devices since it can minimize these oscillations.    
\end{abstract}

\pacs{72.20.Pa, 73.21.La, 71.15.Mb}

\keywords{Graphene; Quantum dot; Kondo}

\thispagestyle{headings}

\maketitle

\section{Introduction}

Single electron devices made up of a confined 
nanostructure located in the vicinity of 
electrodes that act as electron baths constitute
a viable alternative to traditional field
effect transistors which are about to reach
their physical minimum size in a few years \cite{Likharev03}.
Therefore, detailed study of these devices has
accelerated recently aided by the tremendeous 
advances in nanotechnology during the past 
two decades. To this end, it is of critical
importance to simulate the switching
characteristics of these prototypes in the 
presence of realistic electron-electron and 
electron-phonon interactions to assess their 
suitability for large electrical circuits.

Early studies largely focused on the influence
of the many-body Kondo resonance on the transient
electrical conductance and discovered that 
varios time scales \cite{PlihaletAl05PRB,IzmaylovetAl06JPCM}
emerge after sudden switching of the gate or bias 
voltage \cite{NordlanderetAl99PRL,PlihaletAl00PRB,MerinoMarston04PRB}. 
These timescales are inherently related to the 
spin fluctuations that govern the amount
of time required for full formation of the
Kondo resonance \cite{NordlanderetAl99PRL}. 
Moreover, it turns out that the Van Hove
singularities in the electrode density of
states can induce dramatic effects in transient
conductance. Damped oscillations arising
from interference between the emerging Kondo 
resonance and the singularities in the contact 
density of states have been initially reported 
using prototypical density of states for an 
asymmetrically coupled system \cite{GokeretAl07JPCM}.
Subsequent studies undertaken to make the theoretical 
predictions more realistic used the electrode 
density of states calculated via ab initio methods
as an input in many body calculations and 
exposed beating phenomena in transient electrical
conductance \cite{GokeretAl10PRB,GokeretAl11CPL}.

Kondo effect is probably the most elegant and exotic
example of many body physics whose signature has been 
originally traced in magnetic impurities embedded
in bulk metals as resistance enhancement at low
temperatures \cite{Kondo64PTP}. It took some four 
decades to unravel its existence in confined 
nanostructures like quantum dots \cite{GoldhaberetAl98Nature} 
and molecules \cite{LiangetAl02Nature,ParketAl02Nature}. 
The underlying physical mechanism for the Kondo 
effect is the spin flip processes taking place 
at low temperatures between an isolated spin 
at dot energy level and the Fermi sea of electrons 
at electrodes. When the dot energy level lies 
below the Fermi level, a sharp resonance 
develops around the Fermi level and it causes
an enhancement in current due to the opening of
an extra transport channel. The linewidth of 
this resonance is on the order of an energy scale
called the Kondo temperature $T_K$ given by
\begin{equation}
T_K \approx \left(\frac{D\Gamma_{tot}}{4}\right)^\frac{1}{2}
\exp\left(-\frac{\pi|\epsilon_{\rm dot}|}{\Gamma_{tot}}\right),
\label{tkondo}
\end{equation}
where $D$ is the half bandwidth of the conduction 
electrons and $\Gamma_{tot}$ is the total coupling 
of the dot to the electrodes. In this paper, we will
take $D=9\Gamma_{tot}$.

Introducing a temperature gradient across the electrodes
provides another intriguing and realistic path for
investigation. The existence of Kondo resonance
can be checked directly via a measurement of the 
thermopower (Seebeck coefficient) $S$ and the 
thermoelectric figure of merit $ZT$ which 
is directly related to the thermopower is a key 
indicator of the device's performance. Some
experiments in molecular junctions have recently 
probed this by adjusting the alignment of the 
impurity orbital with respect to the Fermi level
of the contacts and this has been found to result
in a change of sign in thermopower 
\cite{Reddyetal07Science,Bahetietal08NL,Malenetal09NL,TangetAl10APL}.   

Behaviour of the thermopower and figure of merit
has been elucidated as a function of temperature
both in Kondo \cite{YangetAl10PLA,DongetAl02JPCM,CostietAl10PRB}
and mixed valence regimes \cite{CostietAl10PRB}.
These studies were largely in line with the
experimental work and confirmed the critical
role the impurity energy level plays in the sign 
of thermopower. Nevertheless, they were by no
means sufficient to shed light on the thermal switching
behaviour of these devices since they were performed
in steady state. Thermal switching will be required
in practical applications to be able to operate
the device at different thermoelectric figure of
merit values. Therefore, it is imperative to
understand the behaviour of the thermal switch
in transient region. First analysis of the thermopower
in a time-dependent situation showed that the 
the inverse of the saturated decay time of 
thermopower to its steady state value is equal 
to the Kondo temperature when Kondo resonance 
emerges as a result of the sudden motion of the 
dot level to a position nearby the Fermi level 
of the electrodes \cite{GokeretAl12PLA}. Later, 
ac driving of the dot level sinusoidally 
revealed complex fluctuations in instantaneous
thermopower whose extreme values are quite
sensitive to the amplitude of the oscillation
\cite{GokeretAl13JPCM}.

These studies utilized parabolic density
of states for contacts and the effect of 
singularities on the instantaneous thermopower 
and figure of merit in Kondo regime remains
to be seen. It is our aim in this paper to 
fill this gap and investigate the behaviour
of instantaneous thermopower and figure of merit
for a quantum dot whose energy level is shifted
to a position close to the Fermi level of contacts.
This gives rise to the gradual appearance of
the Kondo resonance slightly above the Fermi
level. 

\section{Theory}    

We will assume that the quantum dot energy levels are 
filled completely until its highest occupied molecular 
(HOMO) level with two electrons of opposite spins. We will 
represent the HOMO level with a single discrete spin 
degenerate energy level $\varepsilon_{dot}$ coupled to 
electrodes which are free electron baths. This physical 
system can be mapped to the Anderson hamiltonian which 
can be written as 
\begin{eqnarray}
H(t)&=&\sum_{\sigma} \varepsilon_{dot}(t)n_{\sigma}
+\sum_{k\alpha\sigma}\varepsilon_{k}n_{k\alpha\sigma}
+{\textstyle\frac{1}{2}}{\sum} U_{\sigma,\sigma'}n_{\sigma}n_{\sigma'}+ \nonumber \\
& &\sum_{k\alpha\sigma} \left[ V_{\alpha}(\varepsilon_{k\alpha},t)c_{k\alpha\sigma}^{\dag}c_{\sigma}+
{\rm H.c.} \right],
\label{Anderson}
\end{eqnarray}
where $c^\dagger_\sigma$ ($c_\sigma$) and 
$c^\dagger_{k\alpha\sigma}$ ($c_{k\alpha\sigma}$) with 
$\alpha$=L,R create (destroy) an electron of spin 
$\sigma$ in the dot energylevel and in the left(L) and 
right(R) electrodes respectively. The $n_\sigma$ and 
$n_{k\alpha\sigma}$ are occupancies of the dot energy 
level and the electrode $\alpha$. $V_{\alpha}$ are 
the tunneling amplitudes between the electrode $\alpha$ 
and the quantum dot. This model is an ideal venue 
to study Kondo correlations since it can be mapped
directly to Kondo model via Schrieffer-Wolf 
transformations. The Coulomb repulsion energy $U$ in
this model represents the charging energy of the 
dot. This corresponds to the energy gap between
the HOMO level and the lowest unoccupied molecular 
(LUMO) level. We take the charging energy to be infinite. 
This is in line with the experiments which indicate 
that the typical charging energy of quantum dots 
overwhelm the other tunable parameters in the Kondo 
regime. This choice results in single occupancy of 
the dot energy level with up or down spin. Moreover, 
we will be using atomic units where we take 
$\hbar=k_{\rm B}=e=1$.

We redefine the electron operators on the dot by 
introducing a massless (slave) boson operator and 
a pseudofermion operator \cite{Coleman84PRB} as
\begin{equation}
c_{\sigma}=b^{\dagger} f_{\sigma}.
\end{equation}
These new operators must also satisfy
\begin{equation}
Q=b^{\dagger}b+\sum_{\sigma}f^{\dagger}_{\sigma}f_{\sigma}=1.
\end{equation}
which ensures one to one mapping between the Anderson
hamiltonian and the transformed hamiltonian in 
$U \rightarrow \infty$ limit by limiting the dot 
energy level occupancy to single electron. We can
safely omit the Hubbard term from the transformed
Hamiltonian which turns out to be 
\begin{eqnarray}
H(t)&=&\sum_{\sigma}\epsilon_{dot}(t)n_{\sigma}+ \nonumber \\
& &\sum_{k\alpha\sigma}\left [\epsilon_{k}n_{k\alpha\sigma}+
V_{\alpha}(\varepsilon_{k\alpha},t)c_{k\alpha\sigma}^{\dag}
b^{\dag}f_{\sigma}+{\rm H.c.} \right],
\label{slave}
\end{eqnarray}
where $f_{\sigma}^{\dag}(f_{\sigma})$ and $b^{\dag}(b)$
with $\alpha$=L,R are creation(destruction) operators 
for an electron with spin $\sigma$ and a slave boson 
on the dot energy level. 

Taking time independent hopping matrix elements, the 
total coupling of the quantum dot to the electrodes 
is given by $\Gamma_{L(R)}(\epsilon)=\bar{\Gamma}_{L(R)} \rho_{L(R)}(\epsilon)$,
where $\bar{\Gamma}_{L(R)}$ is a constant given by
$\bar{\Gamma}_{L(R)}=2\pi|V_{L(R)}(\epsilon_f)|^2$ 
and $\rho_{L(R)}(\epsilon)$ is the density of states 
function of the electrodes. Both left and right 
electrodes are taken to be made up of the same 
material in conjunction with experiments implying that 
$\rho_{L}(\epsilon)$=$\rho_{R}(\epsilon)$=$\rho(\epsilon)$.
Consequently, we obtain
$\Gamma_{tot}(\epsilon)=\bar{\Gamma}_{tot}\rho(\epsilon)$
where $\Gamma_{tot}(\epsilon)=\Gamma_{L}(\epsilon)+\Gamma_{R}(\epsilon)$
and $\bar{\Gamma}_{tot}=\bar{\Gamma}_{L}+\bar{\Gamma}_{R}$.
  
\begin{figure}[htb]
\centerline{\includegraphics[angle=0,width=9.4cm,height=6.8cm]{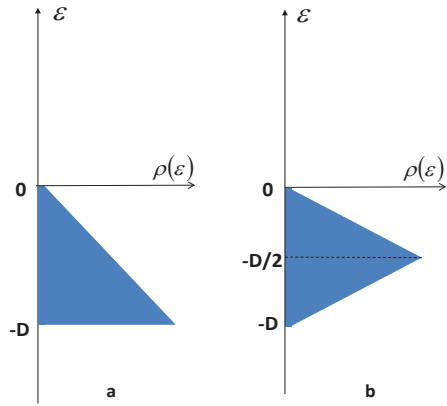}}
\caption{
This figure illustrates the density of states function
$\rho(\epsilon)$ of contacts to which the quantum
dot is attached. Since both left and right electrodes
are identical, only one of them is shown. 
}
\label{Schematic}
\end{figure}

Graphene is a a highly resilient and versatile 
material which has been manufactured in large 
quantities quite recently thanks to the scotch 
tape method \cite{GeimetAl07NM}. Its unique features 
largely stem from the fact that it can be considered 
both a semiconductor with zero bandgap and a metal 
whose density of states is zero at the Fermi level
\cite{CastroNetoetal09RMP}. The first type of contact 
we choose in Fig.~\ref{Schematic} mimics the density 
of states of graphene $\rho(\epsilon)$ except that 
we added an infinitesimal thick cylindirical feature 
inside the conic structure which results in nonzero 
value at the Fermi level to enable the proper development 
of the Kondo resonance. In this case, the Van Hove 
singularity lies at the bottom of the band. Since we
assume that the dot is weakly coupled to the electrodes,
the band cutoff at $D=9\Gamma_{tot}$ lies at a sufficiently
low energy. This enables us to safely ignore deviations
from the linear dispersion relation at high energies and
the Dirac cone approximation is still valid. 
The second type of contact depicted in Fig.~\ref{Schematic}
is a prototypical metal and of triangular shape. Once 
again we added a cylindirical feature with infinitesimal 
thickness inside the triangle to enable nonzero value 
at the Fermi level. The Van Hove singularity lies in 
the middle of the band in this latter case. 

In both cases, we showed only the valence bands 
in Fig.~\ref{Schematic} for simplicity. The conduction 
bands are just the mirror image of valence bands across
the Fermi level. Furthermore, we made sure the number of 
electrons is the same in both cases by integrating the 
area from the bottom of the band to the top. The 
latter one is a necessary step to facilitate direct 
comparison between the two structures. In particular, 
the density of states function for graphene in 
Fig.~\ref{Schematic}a is given by
\begin{eqnarray}
\rho(\epsilon)=2.17\mid\epsilon\mid+\delta &if& \mid\epsilon\mid \le D \nonumber \\
\rho(\epsilon)=0 &if& \mid\epsilon\mid > D,
\end{eqnarray}
where the prefactor corresponds to the inverse
of the Fermi velocity in atomic units and $\delta$ 
is an infinitesimal number. On the other hand, 
the density of states function for the metal 
shown in Fig.~\ref{Schematic}b is 
\begin{eqnarray}
\rho(\epsilon)=4.34\mid\epsilon\mid+\delta &if&  \mid\epsilon\mid \le 0.5D \nonumber \\
\rho(\epsilon)=-4.34\mid\epsilon\mid+\delta+4.34D &if& 0.5D \le \mid\epsilon\mid \le D \nonumber \\
\rho(\epsilon)=0 &if& \mid\epsilon\mid> D.  
\end{eqnarray}

Our technique to tackle the slave boson Hamiltonian
in Eq.~(\ref{slave}) relies on the double time Kadanoff-Baym 
Green functions which have been adapted to the quantum 
impurity problems \cite{LangrethNordlander91PRB,ShaoetAl194PRB}. 
In particular, the retarded Green function is given by
\begin{equation}
G^R(t,t_1) = -i\theta(t-t_1)<\{c_{\sigma}(t),c^{\dagger}_{\sigma}(t_1)\} >,
\end{equation}
where the curly brackets denote anticommutation. 
After performing the slave boson transformation 
described above, the retarded Green function can 
be redefined in terms of the slave boson and pseudofermion
Green functions \cite{GokeretAl07JPCM} as 
\begin{eqnarray}
G^R(t,t_1) &=& -i\theta(t-t_1)[G^R_{pseudo}(t,t_1)B^<(t_1,t) \nonumber \\
& & +G^<_{pseudo}(t,t_1)B^{R}(t_1,t)].
\end{eqnarray}

The task then boils down to calculating these
double time Green functions. We handle this
by solving coupled Dyson equations in a two-dimensional
cartesian grid whose dimensions correspond to
the time arguments of the Green function. Care
must be taken in this procedure to build the 
memory effect into the calculation. This is
accomplished by performing a convergence test
via increasing size of the matrix in which the 
values of the Green function are stored. 

In order to solve the coupled Dyson equations,
the pseudofermion and slave boson self energies
must be approximated somehow since they are 
critical inputs to these equations. We omit 
higher order corrections and thus are content
with the non-crossing approximation(NCA)
\cite{ShaoetAl194PRB,IzmaylovetAl06JPCM} 
for this purpose. As crude as it looks, it 
is known to give remarkably correct results
for dynamical quantities at zero magnetic 
field outside the Fermi liquid regime which 
corresponds to temperatures much smaller than 
the Kondo energy scale. We will stay away from 
the Fermi liquid regime in this paper for this reason. 
Once the Green functions are obtained in this method,
the matrix that stores their values is propagated 
in time in discreet steps diagonally. This 
is how the instantaneous values of the physical 
quantities are determined.

The physical quantities we will be interested
are electrical conductance $G(t)$, thermal
conductance $\kappa(t)$ and thermopower $S(t)$.  
In linear response, their instantaneous values
can be expressed in terms of the Onsager 
coefficients given by
\begin{eqnarray}
& & L_{11} (t)= T \times \nonumber \\
& & Im \left(\int_{-\infty} ^t dt_1 \int \frac{d\epsilon}{2\pi} e^{i\epsilon(t-t_1)} \Gamma_{tot}(\epsilon) G^r (t,t_1) \frac{\partial f(\epsilon)} {\partial \epsilon}\right), \nonumber \\
& &L_{12} (t)= T^2 \times \nonumber \\
& & Im \left(\int_{-\infty} ^t dt_1 \int \frac{d\epsilon}{2\pi} e^{i\epsilon(t-t_1)} \Gamma_{tot}(\epsilon) G^r (t,t_1) \frac{\partial f(\epsilon)} {\partial T}\right), \nonumber \\
& &L_{22} (t)= T^2 \times \nonumber \\
& & Im \left(\int_{-\infty} ^t dt_1 \int \frac{d\epsilon}{2\pi} e^{i\epsilon(t-t_1)} \Gamma_{tot}(\epsilon) \epsilon  G^r (t,t_1) \frac{\partial f(\epsilon)} {\partial T}\right),
\label{Onsager}
\end{eqnarray}
where $f(\epsilon)$ is the Fermi-Dirac distribution function 
given by $f(\epsilon)=\frac{1}{1+e^{\beta \epsilon}}$
with $\beta=\frac{1}{T}$.

In terms of these coefficients, the physical quantities 
can be cast as 
\begin{eqnarray}
G(t)&=&\frac{L_{11}(t)}{T},  \nonumber \\
\kappa(t)&=& \frac{1}{T^2}\left (L_{22}(t)-\frac{L_{12}^2 (t)}{L_{11} (t)} \right), \nonumber \\
S(t)&=& \frac{L_{12}(t)}{T L_{11} (t)}.
\label{definition}
\end{eqnarray}
On the other hand, the figure of merit of a nanodevice
is a dimensionless quantity which is an indicator of
its efficiency. Hence, it is a benchmark on its 
performance in a circuit. In terms of the previously
defined quantities,it is given by
\begin{equation}
ZT=\frac{S^2(t) G(t) T}{\kappa(t)}.
\label{merit}
\end{equation}
The figure of merit always acquires a positive value
and larger figure of merit values imply better 
performance. We will be ignoring the phonon contribution
to the thermal conductance. This assumption is justified
because we will be carrying out our calculations around
the Kondo temperature of the device which is only a few 
Kelvins. Phonon contribution to the thermal conductance 
can be safely ignored in this regime compared to the electronic
one in contrast with some previous work where the system is
outside Kondo regime and the ambient temperature is 
several hundred Kelvins 
\cite{BergfieldetAl10ACSNano,TsaousidouetAl10JPCM}.

In this paper, we will be concerned with the behaviour of
instantaneous thermopower and figure of merit for a 
quantum dot whose energy level is shifted suddenly
to a position close to Fermi levels of contacts 
such that the Kondo resonance can develop in
this final state. We will have the chance to 
make a direct comparison between the effect of
metal and graphene type electrodes on transient 
values of these quantities. This constitutes 
a fairly realistic model to simulate the thermal
switching behaviour of single electron devices
since electron-electron interactions are captured
as well as the structure of the electrode material.
Hence, our investigation should shed light on 
design and implementation of these devices 
experimentally. Our only limitation will be
that our results will pertain to linear response
only due to the validity of Onsager coefficients
in this regime. This is also realistic because
infinitesimal temperature gradient between the
electrodes can be achieved easily by illuminating
one of the contacts with a faint laser beam.

\section{Results and Discussion}

In this section, we will present our numerical
results on the instantaneous thermopower and 
figure of merit of a single quantum dot whose 
energy level is abruptly moved from 
$\epsilon_{dot}$=-5$\Gamma_{tot}$ to
$\epsilon_{dot}$=-2$\Gamma_{tot}$ via a gate 
voltage. This motion triggers a transition 
from a state where the Kondo resonance is 
completely absent to a state where the
Kondo resonance gradually emerges. This
transition takes place due to a drastic
change in Kondo temperature $T_K$ since 
it critically depends on the position of
the dot energy level $\epsilon_{dot}$
as seen in Eq.~(\ref{tkondo}).  

\begin{figure}[htb]
\centerline{\includegraphics[angle=0,width=8.7cm,height=6.0cm]{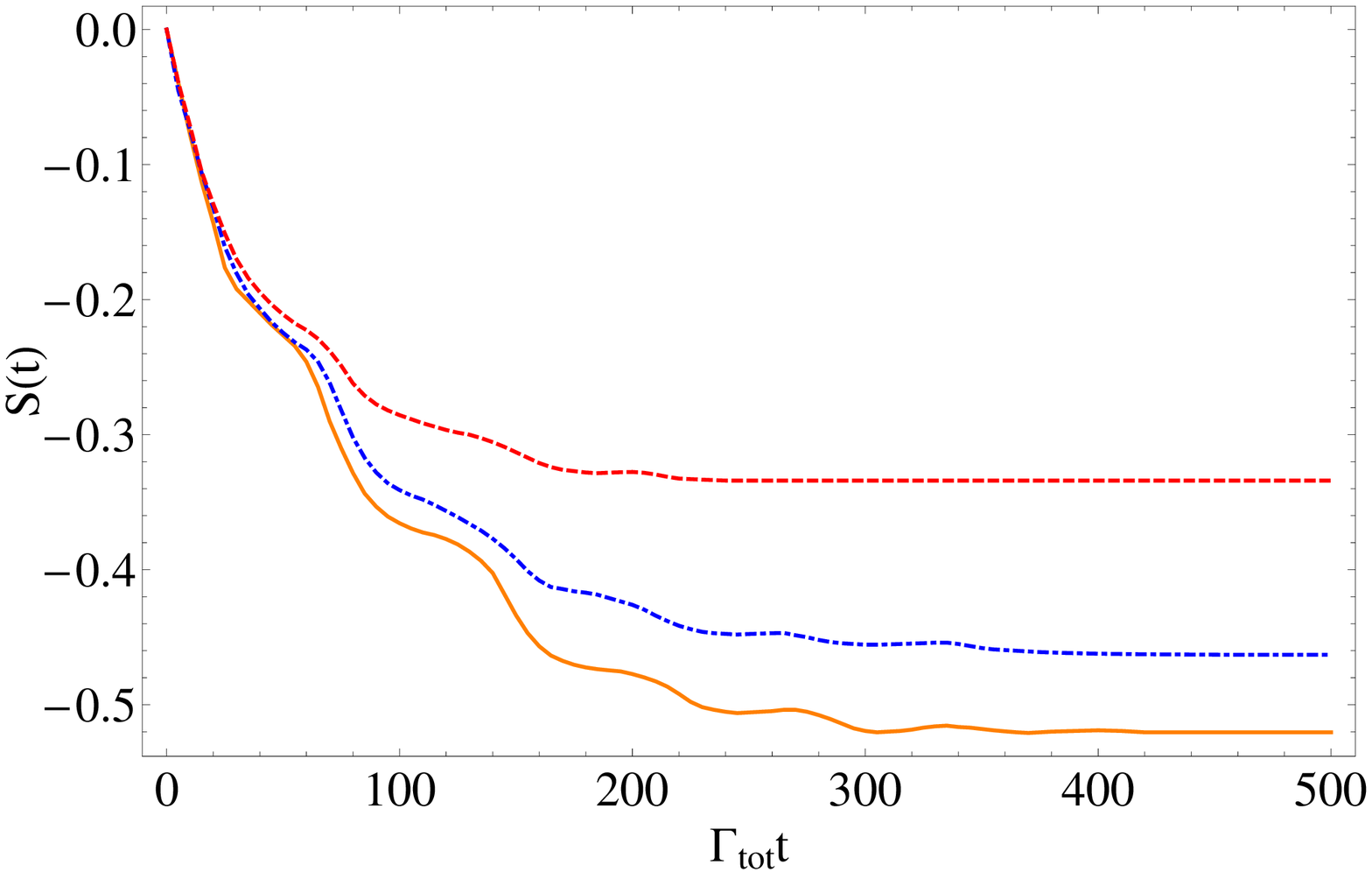}}
\caption{
This figure shows the instantaneous thermopower
$S(t)$ immediately after the dot energy level
has been moved to its final position
for ambient temperatures of
$0.0035\Gamma_{tot}$ (red dashed),
$0.0028\Gamma_{tot}$ (blue dot dashed) and
$0.0014\Gamma_{tot}$ (orange solid)
in linear response for metal   
electrodes shown in Fig.~\ref{Schematic}b.
}
\label{Fig2}
\end{figure}

We will start our analysis with the instantaneous 
thermopower results. Fig.~\ref{Fig2} depicts the 
instantaneous thermopower for three different ambient
temperatures immediately after the dot level has 
been switched to its final location for metal 
electrodes shown in Fig.~\ref{Schematic}b. The
Kondo temperature in the final state is 
inferred to be $T_K$=0.0021$\Gamma_{tot}$.
Consequently, the higher two temperatures 
are above $T_K$ while the lowest temperature
is slightly lower. We see in this figure that
damped sinusoidal oscillations start developing 
at the onset of Kondo timescale with the same 
frequency for all temperatures.

Kondo timescale, during which the Kondo resonance 
starts emerging slightly above the Fermi level,
takes place after the short timescale which 
corresponds to charge fluctuations. Short timescale
roughly takes place between 0$<\Gamma_{tot}t<$10.
This result is quite interesting and different from 
conductance oscillations which appear much later
at $\Gamma_{tot}t>$80 \cite{PlihaletAl05PRB}.
Another novelty is the insensitivity of these
oscillations to asymmetry of the junction 
defined as $\eta=\bar{\Gamma}_{L}/\bar{\Gamma}_{tot}$.
Changing the asymmetry factor $\eta$ does not 
effect the frequency and amplitude of the 
oscillations. Moreover, the final steady 
state thermopower (i.e. $\Gamma_{tot}t \rightarrow \infty$)
also remains unchanged when $\eta$ is altered in 
conjunction with other studies investigating steady 
state thermopower\cite{KrawiecetAl07PRB}. The 
only resemblance these results bear to 
instantaneous conductance data is the saturation 
of the amplitudes below the Kondo temperature. 
We did not show the curves well below $T_K$ in 
Fig.~\ref{Fig2} as they overlap with higher 
temperature ones and obscure the oscillations.
The oscillation frequency is the same at all
temperatures and the oscillations persist 
deep into the long timescale with $\Gamma_{tot}t>$300
when $T \le T_K$.

\begin{figure}[htb]
\centerline{\includegraphics[angle=0,width=8.7cm,height=6.0cm]{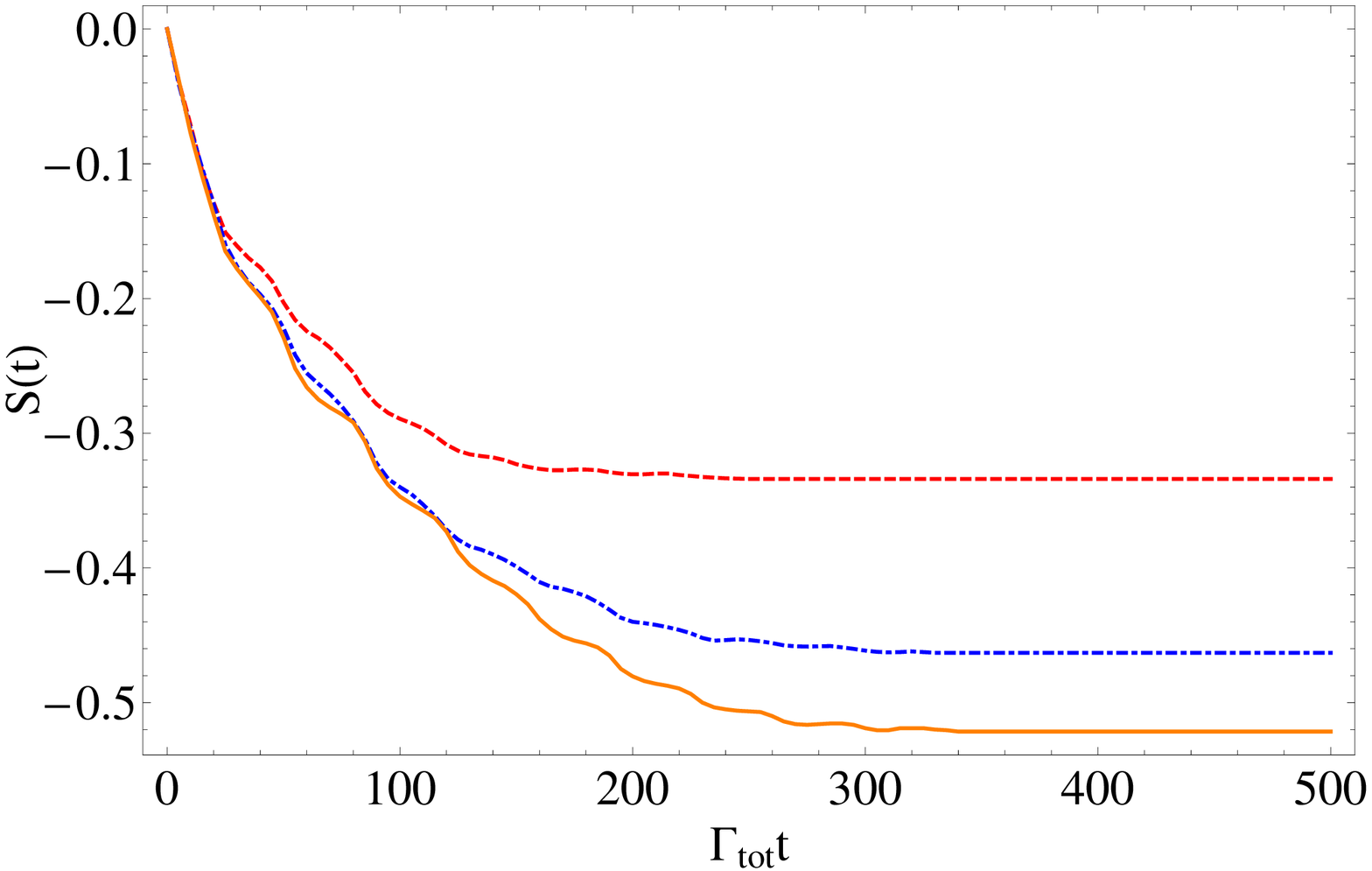}}
\caption{
This figure shows the instantaneous thermopower
$S(t)$ immediately after the dot energy level 
has been moved to its final position
for ambient temperatures of               
$0.0035\Gamma_{tot}$ (red dashed),
$0.0028\Gamma_{tot}$ (blue dot dashed) and
$0.0014\Gamma_{tot}$ (orange solid)
in linear response for graphene 
electrodes shown in Fig.~\ref{Schematic}a.
}
\label{Fig3}
\end{figure}

Fig.~\ref{Fig3} displays our numerical 
results for the instantaneous thermopower 
using the same parameters used in Fig.~\ref{Fig2}.
The major difference is that the quantum dot
is attached to graphene electrodes shown in
Fig.~\ref{Schematic}a. The oscillations starting
in the Kondo timescale are still present here.
However, their amplitudes are suppressed dramatically
compared to metal electrodes. On the other hand,
the steady state thermopower values are still the 
same as Fig.~\ref{Fig2} but the oscillation frequency
doubled. These results obviously suggest that 
graphene is a more desirable material to be used 
as electrode in single electron devices because 
it enables to minimize if not completely suppress 
the undesirable fluctuations during the switching 
of these devices. This is a great advantage in 
integrating these devices into larger circuits.  
  
\begin{figure}[htb]
\centerline{\includegraphics[angle=0,width=8.7cm,height=6.0cm]{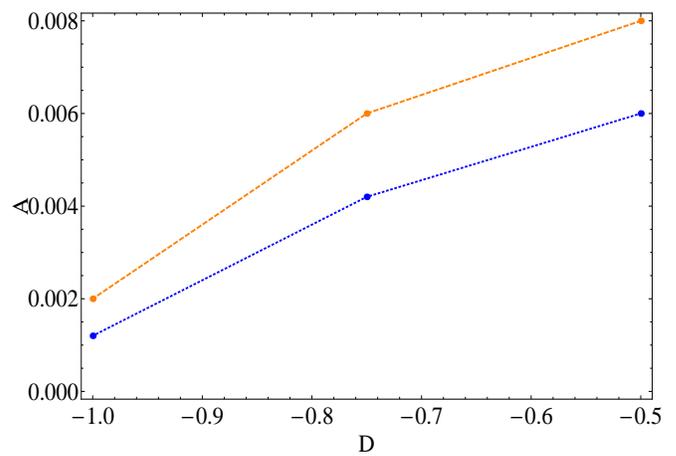}}
\caption{
This figure shows the amplitude of the 
peak or trough nearest to $\Gamma_{tot}t$=200
that occurs in thermopower oscillations
at ambient temperature of 
$0.0014\Gamma_{tot}$ (orange dashed)
and $0.0028\Gamma_{tot}$ (blue dotted) 
as a function of the position of the 
van Hove singularity in the contact
density of states. Solid dots represent
the actual data points.
}
\label{Fig4}
\end{figure}

We want to dwell on the underlying microscopic
mechanism for these oscillations. The numerical 
results presented above show that the frequency of
oscillations is proportional to the energy separation
between the Fermi level of the contacts and the sharp feature
in the density of states of the contacts. That feature
happens to be in the edge of the band for graphene 
therefore it generates the highest frequency. More
importantly, the amplitude of the oscillations decreases
as the energy separation between the Fermi level and 
the sharp feature increases resulting in the minimum
amplitude for graphene. 

In order to generalize and verify this conclusion, we 
performed additional calculations with the same parameters 
in Fig.~\ref{Fig2} and Fig.~\ref{Fig3} by using metal electrodes 
whose density of states is triangular like Fig.~\ref{Schematic}b
except the fact that the van Hove singularity is located at 
-0.75$D$. We calculated the amplitude of the nearest peak 
or trough occurring around $\Gamma_{tot}t$=200 for all 
three electrodes. The amplitude has been determined by 
taking the average of the two amplitudes corresponding 
to the two adjacent peak-trough pairs. This step is 
necessary because the oscillations are damped. The results 
are shown in Fig.~\ref{Fig4} for two ambient temperatures
as a function of the position of the van Hove singularity
and it is obvious from here that the graphene electrodes 
where the van Hove singularity is at -$D$ produce
minimum amplitude at both temperatures. The instant
$\Gamma_{tot}t$=200 is arbitrary and same calculation
at other instants yields similar results. If the dot
gets attached to electrodes made up of two dimensional 
electron gas with flat density of states and the same 
bandwidth $D$, the oscillation frequency is the same 
as graphene because the van Hove singularity still lies
at the edge of the band. However, the oscillation amplitudes
turn out to be larger than graphene presumably due to 
the large difference in the value of the density of 
states at Fermi level.

\begin{figure}[htb]
\centerline{\includegraphics[angle=0,width=8.7cm,height=6.0cm]{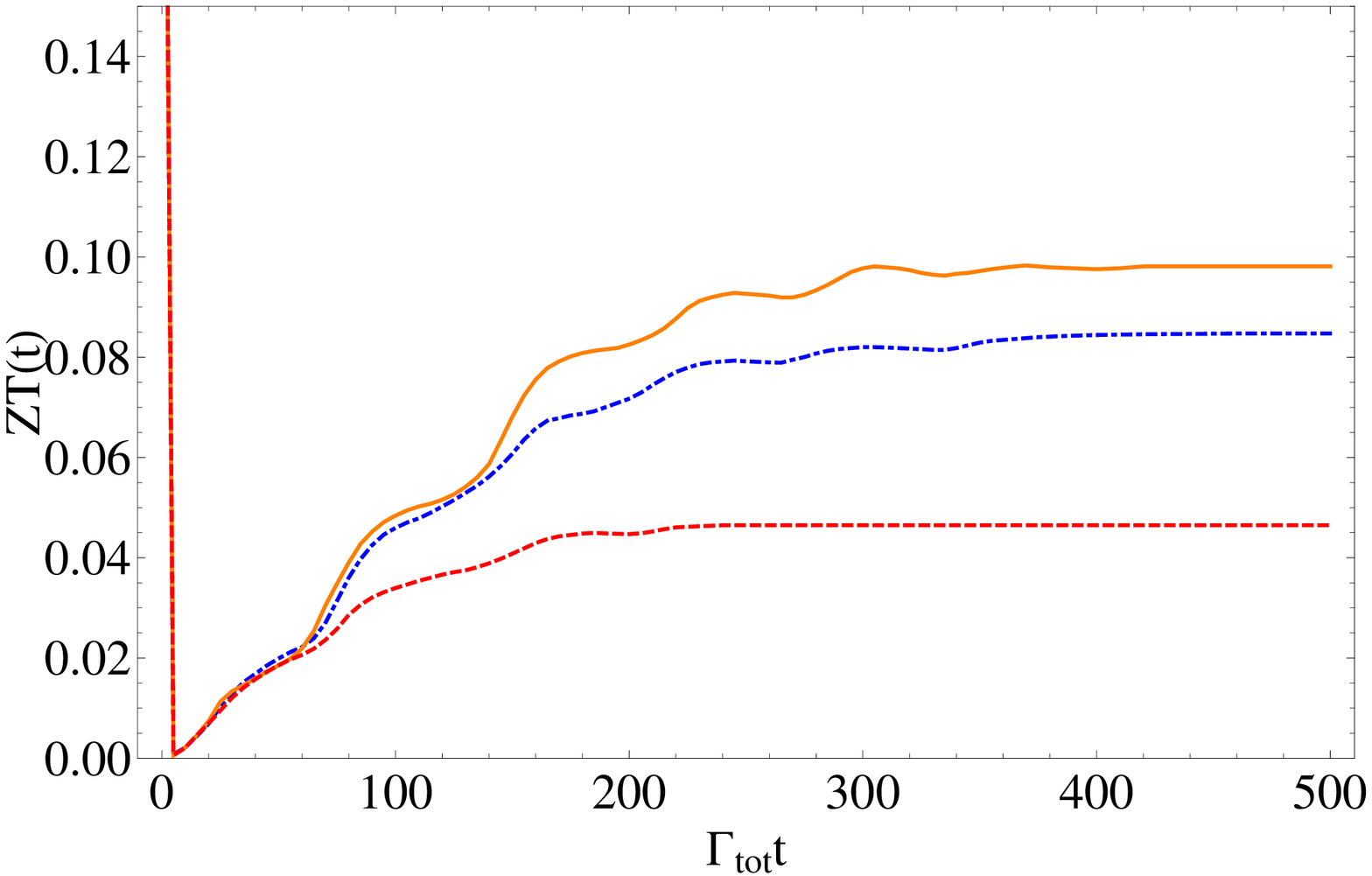}}
\caption{
This figure shows the instantaneous figure of merit
$ZT(t)$ immediately after the dot energy level       
has been moved to its final position
for ambient temperatures of
$0.0035\Gamma_{tot}$ (red dashed),                
$0.0028\Gamma_{tot}$ (blue dot dashed) and        
$0.0014\Gamma_{tot}$ (orange solid)
in linear response for metal
electrodes shown in Fig.~\ref{Schematic}b.          
}
\label{Fig5}
\end{figure}

These results suggest the presence of an interference
process. The components of this interference should be 
the two sharp features. These are nothing but the
van Hove singularity in the density of states 
of the electrodes and the Kondo resonance in the dot 
density of states. This interference is enabled due to
two factors. First, the density of states of the dot
and the electrodes are mixed due to the hybridization
term in the Hamiltonian therefore the Kondo resonance
and the van Hove singularity can interact even though
they are not located in the same density of states. Second,
the development of the Kondo resonance is gradual and it
is fully formed only at the end of the long timescale
\cite{PlihaletAl05PRB} implying that it is a dynamical 
time-dependent feature. On the other hand, the 
van Hove singularity at the contact density of states 
is static and it does not evolve in time. Consequently,
the interference can take place between a dynamic and a static
feature. It dies out once they both become static, hence
we observe damped oscillations.

The thermopower is related to the dot density of states 
at low temperatures via Sommerfeld expansion which is given by
\begin{equation}
S(T)=-\frac{\pi^2 T}{3 A(0,T)}\frac{\partial A}{\partial \epsilon} \rvert_{\epsilon=0}
\end{equation}
in atomic units. $A(0,T)$ corresponds to the value 
of the dot density of states at Fermi level of the 
contacts and $\frac{\partial A}{\partial \epsilon}$
is its derivative. The formation of the Kondo resonance 
has been previously shown to take place slightly above 
the Fermi level \cite{CostietAl94JPCM} giving rise to 
a positive derivative at Fermi level. Consequently, the 
overall sign of the thermopower in the final state 
is always negative. As the Kondo resonance starts 
growing in the final state, the derivative begins 
to increase in magnitude. When the Kondo resonance 
reaches its full form in the long timescale, the 
derivative stabilizes. This causes the thermopower 
to reach its steady state. 

The fundamental difference between the oscillations 
in the thermopower and the conductance reported previously 
lies in their start of occurrence and sensitivity to 
asymmetry. The thermopower oscillations start much 
earlier than the conductance oscillations because tiny
fluctuations in the dot density of states at Fermi
level due to this interference are enhanced as a 
result of the derivative in Sommerfeld expansion and 
thus manifest themselves in the thermopower at the 
onset of Kondo timescale prominently. The insensitivity 
to asymmetry originates from the constructive interference 
of the right and left electrodes with the Kondo resonance 
magnifying the effect even further. The destructive 
interference between the left and right electrodes 
diminishes the conductance oscillations gradually 
with increasing $\eta$ 
\cite{GokeretAl07JPCM,GokeretAl10PRB,GokeretAl11CPL}.

We also want to go over the instantaneous figure 
of merit results. We use the same parameters in 
Fig.~\ref{Fig2} and Fig.~\ref{Fig3} for this
purpose. Fig.~\ref{Fig5} displays the instantaneous 
figure of merit immediately after the dot level 
has been switched to its final position for metal 
electrodes. A first glance at this figure
reveals that the figure of merit is nonzero
in initial state on contrary to the thermopower
shown in Fig.~\ref{Fig2} and Fig.~\ref{Fig3}.
This can be understood by simplifying the 
figure of merit using the Wiedemann-Franz law.
The ratio of the thermal conductivity to 
electrical conductivity is given by the 
Wiedemann-Franz law which can be expressed as 
\begin{equation}
\frac{\kappa(t)}{G(t)}=LT,
\label{wf}
\end{equation}
where $L$ is a constant called the Lorentz number.
It can be cast as $L=\pi^2/3$ in atomic units.
Inserting this identity into Eq.~(\ref{merit})
gives 
\begin{equation}
ZT=S^2(t)/L.
\label{ZTnew}
\end{equation}

This obviously implies that the thermoelectric
figure of merit can never be negative and is
proportional to the square of the instantaneous
thermopower which we just investigated. However,
the main subtlety is that the value of the Lorentz 
number deviates from its real value in mesoscopic 
systems including quantum dots outside the Fermi 
liquid state. Fermi liquid state occurs when $T \ll T_K$. 
Incidentally, our calculations invoking NCA is valid outside 
the Fermi liquid state too. In the initial dot energy
level, $T \gg T_K$ and our numerical results reveal a much 
larger thermoelectric figure of merit than the 
final dot energy level owing to the suppression of the 
Lorentz number greatly in line with previous steady 
state results \cite{YangetAl10PLA,DongetAl02JPCM,CostietAl10PRB}. 
Consequently, the instantaneous figure of merit 
exhibits a sharp jump upon switching the dot level 
abruptly at t=0.

\begin{figure}[htb]
\centerline{\includegraphics[angle=0,width=8.7cm,height=6.0cm]{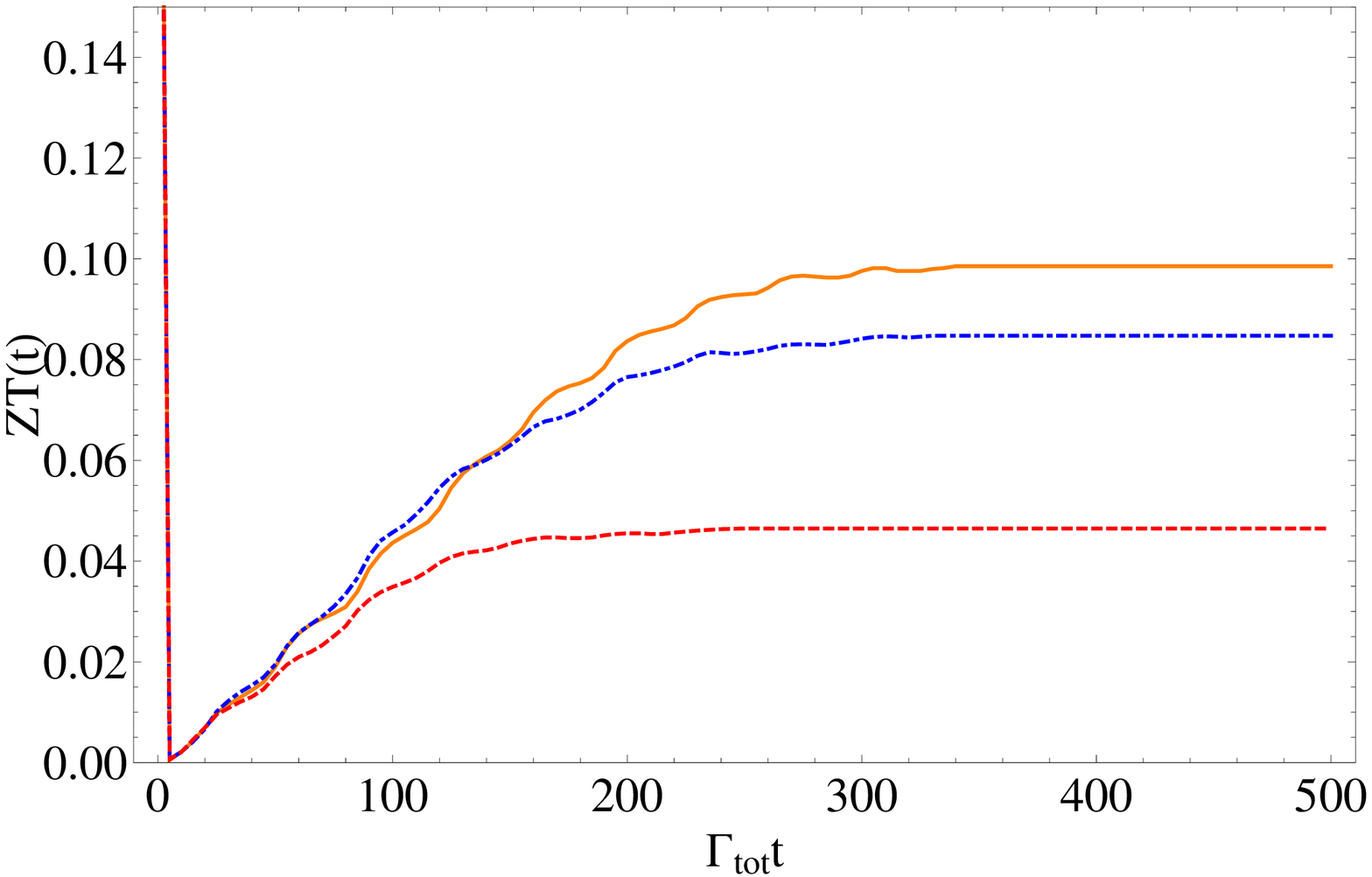}}
\caption{
This figure shows the instantaneous figure of merit
$ZT(t)$ immediately after the dot energy level
has been moved to its final position
for ambient temperatures of
$0.0035\Gamma_{tot}$ (red dashed),
$0.0028\Gamma_{tot}$ (blue dot dashed) and
$0.0014\Gamma_{tot}$ (orange solid)
in linear response for graphene
electrodes shown in Fig.~\ref{Schematic}a.
}
\label{Fig6}
\end{figure}

The behaviour of the instantaneous figure of merit
curves in the final state generally follows the
square of the respective instantaneous thermopower
curves. The oscillation frequency is again the same
in this case. Besides these obvious points, there 
is a much more intricate issue that is worth pointing 
out here. We deliberately chose ambient temperatures 
that are around $T_K$ for our calculations. An 
earlier investigation showed that the value of the
Lorentz number declines with a sharp slope upon
passing through $T_K$ as a function of temperature
\cite{DongetAl02JPCM}. This rapid change of the value
of the Lorentz number plays an important for our time
dependent calculations. This role stems from the fact
that the instantaneous figure of merit is proportional
to the inverse of the Lorentz number as seen
in Eq.~(\ref{ZTnew}).

The $1/L$ serves as an enhancement factor for the 
instantaneous figure of merit. The amplitudes of the 
oscillations taking place at or above the Kondo temperature 
$T_K$ are enhanced more than the amplitudes of the 
oscillations occurring at ambient temperatures below 
$T_K$ as a result of this slope around $T_K$. This 
can be seen in Fig.~\ref{Fig5}. On the other hand, 
Fig.~\ref{Fig6} shows the instantaneous figure of 
merit results for the same parameters used in Fig.~\ref{Fig5}
when the quantum dot is attached to graphene electrodes. 
We see in this figure that the graphene electrodes 
largely suppress the amplitude of the oscillations 
compared to Fig.~\ref{Fig5} due to the larger energy
gap between the Van Hove singularity at the edge of 
the band and the Fermi level. This also doubles the 
oscillation frequency compared to Fig.~\ref{Fig5}.
Main advantage of this suppression is the prevention
of the sudden drop of the device performance at trough
points. 
 
Another interesting peculiarity occurring in Fig.~\ref{Fig6} 
due to the sharp variation of the value of the Lorentz number
across $T_K$ is that the value of the instantaneous 
thermopower at an ambient temperature above $T_K$ can 
temporarily exceed the value of the instantaneous 
thermopower below $T_K$. This can be seen between 
40$<\Gamma_{tot}t<$140. This is clearly a transient 
phenomena and cannot be accessed with a steady-state formalism.

\section{Conclusions}

In conclusion, we studied in linear response both 
the instantaneous thermopower and figure of merit 
for a quantum dot whose energy level is suddenly 
moved into a position such that the many body Kondo 
resonance develops in the final position. We modeled 
graphene and metal type contacts for this purpose. 
We uncovered damped sinusoidal oscillations starting 
at the onset of the Kondo timescale and persisting 
until the long timescale regardless of the asymmetry 
factor. This behaviour is distinct and different from 
the instantaneous conductance oscillations which were 
investigated previously in great detail. The 
frequency of these oscillations turns out to be
proportional to the separation between the Fermi 
level of the contacts and the van Hove singularity 
in the contact density of states. The amplitudes
more or less saturate around the Kondo timescale
suggesting that the interference between the Kondo
resonance and the van Hove singularities is the
underlying physical mechanism for these oscillations.
This interference gives rise to small fluctuations in 
the dot density of states at Fermi level whose effects 
are amplified in the thermopower through its derivative 
in the Sommerfeld expansion. This results in fairly 
large oscillation amplitudes for metallic electrodes. 
However, graphene contacts generate the smallest 
amplitude since their van Hove singularity lies 
farthest away from the Fermi level minimizing the 
interference. This renders graphene as an ideal 
electrode candidate for future single electron devices.

We also investigated the instantaneous figure of
merit for the same set up. Violation of the
Wiedeman-Franz law causes peculiarities here. 
A sharp decline of the value of the Lorentz
number around the Kondo temperature results
in an enhancement of the oscillations above
the Kondo temperature more than the ones 
occurring below the Kondo temperature. Due
to this subtlety, the value of the instantaneous 
thermopower at an ambient temperature above $T_K$ can
temporarily exceed the value of the instantaneous
thermopower below $T_K$ in the Kondo timescale.

As a final note, we would like to comment on
the experimental feasibility of our work. Graphene
sheets can be manufactured routinely via scotch 
tape method and quantum dots are produced via 
molecular beam epitaxy quite easily. The small 
temperature gradient between the contacts can be 
obtained by illuminating one of the contacts with a 
faint laser beam. Moreover, recent remarkable 
progress in ultrafast pump-probe methods 
\cite{Teradaetal10JPCM,TeradaetAl10Nature}
should make measuring thermopower and thermal
conductance in real time possible. 
  
\section{Acknowledgments}

The authors thank T$\ddot{u}$bitak for
generous financial support via grant 111T303.

\bibliographystyle{iopams}
\providecommand{\newblock}{}

\end{document}